\documentclass[reqno,11pt]{amsart}

\usepackage{amssymb, epsf}

\def\be{\begin{equation}}
\def\ba{\begin{align}}
\def\bm{\begin{multline}}
\def\bfig{\begin{figure}[htb]}
\def\efig{\end{figure}}

\newcommand{\bibit}[1]{\vspace{1mm} \bibitem[#1]{#1}}
\newcommand{\paper}[1]{{\it #1}, }
\newcommand{\journal}[4]{#1 {\bf #2}, #3 (#4)}
\newcommand{\CMP}{Commun. Math. Phys.}
\newcommand{\HPA}{Helv. Phys. Acta}
\newcommand{\JSP}{J. Stat. Phys.}
\newcommand{\PR}{Phys. Rev.}

\newcommand{\PRB}{Phys. Rev. B}

\newcommand{\RMP}{Rev. Math. Phys.}

\numberwithin{equation}{section}
\newtheorem{theorem}{Theorem}[section]
\newtheorem{proposition}[theorem]{Proposition}

\newcommand{\fig}{Fig.\;}

\newcommand{\eg}{e.g.\;}
\newcommand{\ie}{i.e.\;}

\newcommand{\nn}{\nonumber}

\renewcommand{\leq}{\;\leqslant\;}
\renewcommand{\geq}{\;\geqslant\;}
\newcommand{\isdefby}{\;\doteqdot\;}

\newcommand{\dd}{{\rm d}}
\newcommand{\e}[1]{\,{\rm e}^{#1}\,}
\newcommand{\ii}{{\rm i}}
\newcommand{\sumtwo}[2]{\sum_{\substack{#1 \\ #2}}}
\newcommand{\sumthree}[3]{\sum_{\substack{#1 \\ #2 \\ #3}}}

\newcommand{\inttwo}[2]{\int_{\substack{#1 \\ #2}}}

\DeclareMathOperator*{\union}{\text{\large$\cup$}}

\def\Tr{{\operatorname{Tr\,}}}

\DeclareMathOperator{\supp}{supp}

\def\supp{{\operatorname{supp\,}}}

\def\bbig{{\operatorname{big\,}}}
\def\short{{\operatorname{short\,}}}

\def\per{{\text{\rm\,per}}}

\def\bra #1{\langle#1 |\,}
\def\ket #1{\,|#1 \rangle}

\newcommand{\expval}[1]{\langle #1 \rangle}

\newcommand{\neighbours}[2]{<\! #1,#2 \! >}

\newcommand{\const}{{\text{\rm const}}}

\newcommand{\Trfunc}{\Phi^{\rm T}}

\def\writefig#1 #2 #3 {\rlap{\kern #1 truecm \raise #2 truecm
\hbox{#3}}}
\def\figtext#1{\smash{\hbox{#1}} \vspace{-5mm}}

\newcommand{\caA}{{\mathcal A}}
\newcommand{\caB}{{\mathcal B}}

\newcommand{\caD}{{\mathcal D}}

\newcommand{\caG}{{\mathcal G}}
\newcommand{\caH}{{\mathcal H}}

\newcommand{\caK}{{\mathcal K}}

\newcommand{\caP}{{\mathcal P}}
\newcommand{\caQ}{{\mathcal Q}}

\newcommand{\caS}{{\mathcal S}}

\newcommand{\bbC}{{\mathbb C}}

\newcommand{\bbL}{{\mathbb L}}

\newcommand{\bbN}{{\mathbb N}}

\newcommand{\bbR}{{\mathbb R}}

\newcommand{\bbZ}{{\mathbb Z}}

\newcommand{\bsA}{{\boldsymbol A}}

\newcommand{\bsmu}{{\boldsymbol \mu}}

\begin{document}

\begin{quote}
\raggedleft
{\small
\JSP\, {\bf 95}, May 1999
}
\end{quote}
\vspace{2mm}

\title{Analyticity in Hubbard models}

\author{Daniel Ueltschi}

\maketitle

\vspace{-5mm}

\begin{centering}
{\small\it
Institut de Physique Th\'eorique\\
\'Ecole Polytechnique F\'ed\'erale de Lausanne\footnote{Present
address: Dept of Mathematics, Rutgers University, 110 Frelinghuysen
Road, Piscataway, New Jersey 08854-8019; ueltschi@math.rutgers.edu;
http://math.rutgers.edu/\~{}ueltschi}\\
}
\end{centering}

\vspace{5mm}

\begin{quote}
{\small
{\bf Abstract.}
The Hubbard model describes a lattice system of quantum particles with local
(on-site) interactions. Its free energy is analytic when $\beta t$ is small, or
$\beta t^2 / U$ is small; here, $\beta$ is the inverse temperature, $U$ the
on-site repulsion and $t$ the hopping coefficient.

For more general models with Hamiltonian $H = V + T$ where $V$ involves local
terms only, the free energy is analytic when $\beta \| T \|$ is small,
irrespectively of $V$. There exists a unique Gibbs state showing
exponential decay of spatial correlations. These properties are
rigorously established in this paper.
}

\vspace{1mm}
\end{quote}

\vspace{3mm}

\section{Introduction}

Electrons in condensed matter feel an external periodic potential due to
the presence of atoms. A natural basis for the Hilbert space describing the
states of the electrons consists in Wannier states, that are indexed by the
sites of the lattice. The Hamiltonian for this Statistical Physics system can
be written in second quantization in terms of Wannier states, and with some
simplifications we obtain a lattice model \cite{Hub}.

Forgetting the initial physical motivation, we can consider a lattice model as
if the particles were really moving on a lattice, and develop a physical
intuition in this case. It may help in understanding the behaviour of the
system.

The most famous lattice model for the description of quantum particles is the
Hubbard model. It consists in a hopping term (discretized Laplacian)
representing the kinetic energy, and a Coulomb interaction between the
particles. This interaction is local, or on-site, meaning that it is expressed
in terms of creation and annihilation operators of a same site. Many interesting
properties of the Hubbard model have been rigorously established, see
\cite{Lieb} for a review; however, the basic questions about magnetism and
superconductivity are still unsolved. The present paper brings a modest
contribution in the sense that interesting phenomena are definitively excluded
for some values of the thermodynamic parameters.

The general setting is as follows. We consider a class of models that include
the Hubbard one, with Hamiltonian
$$
H^\bsmu = V^\bsmu + T^\bsmu .
$$
The vector $\bsmu$ represents a finite number of parameters, such as chemical
potential, magnetic field, ... $V^\bsmu = \sum_{x \in \bbZ^\nu} V^\bsmu_x$ is a
sum of local operators, and $T^\bsmu = \sum_{A \subset \bbZ^\nu} T^\bsmu_A$ is
a finite-range or exponentially decaying quantum ``interaction". The free
energy is shown to be analytic in $\bsmu$ and $\beta$ in the domain
\be
\label{domain}
\beta \sum_{A \ni x} \| T^\bsmu_A \| \e{c |A|} < \const
\end{equation}
where $c$ is a constant depending on the lattice and on the dimension of the
local Hilbert space. What is quite surprising is that the domain does not
depend on the local interaction $V$. The reason is that when $T$ is small with
respect to $\beta$, the sites of the lattice are almost independent, and the
(mean) free energy is essentially that of a model with only one site. Such a
zero-dimensional system is free from phase transitions, hence its free energy
is analytic. High temperature expansions would yield comparable results;
however they do not only require the condition \eqref{domain}, but also $\beta
\| V \| < \const$.

Concerning the Hubbard model, it is not only true that we have analyticity
however strong is the repulsive potential; the latter favours this phase, \ie
the stronger the interactions, the larger the domain. This last result holds at
half filling, and if the ratio $t/U$ is small enough. In fact, at half
filling the
Hubbard model is unitarily equivalent to a series in powers of $t/U$
that starts with $V$ and the Heisenberg model \cite{KS, CSO, MGY}:
$$
H_{\text{Hubbard}} \simeq V + \tfrac{t^2}U H_{\text{Heisenberg}} +
O(\tfrac{t^4}{U^3})
$$
(a rigorous statement can be found in \cite{DFF}). Such a model enters our class, with
$\| T \| \sim \frac{t^2}U$, hence the condition $\beta t^2 /U < \const$.

Another example is the Falicov-Kimball model \cite{GM}; it is a Hubbard model
where only particles of a given spin have hopping, the others being considered
as heavy, static classical particles. The statements for the Hubbard model are
also valid in this case, and were proven by Kennedy and Lieb \cite{KL}.

Section 2 contains precise definitions, statements and proofs for general
systems with on-site interactions. Section 3 is devoted to the Hubbard model;
domains of parameters where analyticity can be rigorously proven are proposed
with explicit bounds, in the case of the 3D square lattice. Finally, the paper
ends with a discussion of the Bose-Hubbard model, for which partial results may
be obtained.

\section{Models with local interactions}

\subsection{General framework}

Let us be more precise and introduce the mathematical framework. Let $\bbL$ a
$\nu$-dimen\-sional lattice; for instance, $\bbL = \bbZ^\nu$, but any other
periodic lattice can be considered. We denote with $\Lambda$ a finite subset of
$\bbL$, and the thermodynamic limit $\lim_{\Lambda \nearrow \bbL} f_\Lambda$
means $\lim_{n \to \infty} f_{\Lambda_n}$ with any sequence of finite volumes
$(\Lambda_n)$ such that $\Lambda_n \subsetneq \Lambda_{n+1}$, and
$\lim_{n \to \infty} |\partial\Lambda_n| / |\Lambda_n| = 0$, where
$\partial\Lambda_n$ is the boundary of $\Lambda_n$. Let $\Omega$ a
finite set with $|\Omega| = S$; we consider the set of ``classical
configurations" $\Omega^\Lambda$. The Hilbert space $\caH_\Lambda$ at finite
volume $\Lambda$ is spanned by the classical configurations, \ie each vector of
$\caH_\Lambda$ is a linear combination of vectors $\ket{n_\Lambda}$, $n_\Lambda
\in \Omega^\Lambda$.

A {\it quantum interaction} $T$ is a collection $(T_A)_{A \subset \bbL}$, where
$T_A$ is a self-adjoint operator with support $A$. Its action is
defined on each $\caH_\Lambda$ with $\Lambda \supset A$, and we have
factorization properties
\be
\bra{n_\Lambda} T_A T_{A'} \ket{n_\Lambda'} = \bra{n_A} T_A \ket{n_A'}
\bra{n_{A'}} T_{A'} \ket{n_{A'}'}
\end{equation}
when $A \cap A' = \emptyset$ (a hopping matrix is an example of a
``quantum interaction''). Let us introduce the connected cardinality $\|
A \|$ of $A \subset \bbL$ as the cardinality of the smallest
connected set containing $A$, \ie
\be
\| A \| = \min_{B \supset A, \, \text{connected}} |B| ;
\end{equation}
notice that $\| A \| = |A|$ when $A$ is connected, and $\| A \| > |A|$ when it
is not. We define the norm of an interaction to be
\be
\label{defnorm}
\| T \|_c = \sup_{x \in \bbL} \sum_{A \ni x} \| T_A \| \e{c \| A \|}
\end{equation}
where $c$ is a positive number. Here, $\| T_A \|$ is the operator norm of $T_A$. We
call $V$ a {\it local interaction} (or {\it on-site interaction}) if $V_A = 0$
for all $|A| \geq 2$; local interactions are denoted by $(V_x)$ instead of
$(V_{\{ x \}})$.

Let $\bsmu \in \bbR^s$ be thermodynamic parameters. The finite volume
Hamiltonian $H^\bsmu_\Lambda$ depends on $\bsmu$ and is given by
\be
H^\bsmu_\Lambda = \sum_{x \in \Lambda} V^\bsmu_x + \sum_{A \subset \Lambda}
T^\bsmu_A .
\end{equation}
We suppose here that both $(V_x^\bsmu)$ and $(T_A^\bsmu)$ are
translation invariant, although periodic interactions could be
considered with only small modifications.
The free energy is given by the limit (whenever it exists)
\be
\label{deffren}
f(\beta, \bsmu) = -\frac1\beta \lim_{\Lambda \nearrow \bbL} \frac1{|\Lambda|}
\log \Tr \e{-\beta H^\bsmu_\Lambda} .
\end{equation}
We write $f_0$ for the ``classical free energy"
\be
f_0(\beta, \bsmu) = -\frac1\beta \log \sum_{n_x \in \Omega} \bra{n_x} \e{-\beta
V^\bsmu_x} \ket{n_x} .
\end{equation}
We notice that $f_0$ is also given by \eqref{deffren} with $T^\bsmu=0$.

A Gibbs state is a functional that attributes to any bounded local operator $K$
the value
\be
\label{defGS}
\expval K = \lim_{\Lambda \nearrow \bbL} \frac{\Tr K \e{-\beta
H_\Lambda}}{\Tr \e{-\beta H_\Lambda}} .
\end{equation}
A Gibbs state is {\it exponentially clustering} if for any two local operators
$K$ and $K'$ there exists $C_{K, K'} < \infty$ (with $C_{K, K'} = C_{t_x K, t_y
K'}$ for any translations $t_x$ and $t_y$, $x,y \in \bbL$) such that
\be
\bigl| \expval{KK'} - \expval K \expval{K'} \bigr| \leq C_{K, K'}
\e{-d(K,K') / \xi}
\end{equation}
for some finite constant $\xi$. Here, $d(K,K')$ is the distance between
supports of $K$ and $K'$.

\subsection{Uniqueness of the Gibbs state}
\label{secunique}

The Hamiltonians we consider possess many symmetries. For instance,
they have translation invariance by assumption; and typical models
have further conserved quantities, such as the total number of
particles, or total spin...

Usually Gibbs states obtained with free boundary conditions
\eqref{defGS}, or with periodic ones, have same symmetry properties than Hamiltonians. To
obtain pure states with symmetry breaking, there are mainly two ways:
to introduce boundary conditions, or to perturb the system.

In the quantum case, boundary conditions may be defined by means of a
suitable {\it boundary interaction} $\partial^\Lambda =
(\partial^\Lambda_A)_{A \subset \bbL}$, where operators
$\partial^\Lambda_A$ are non zero only for subsets $A$ that touch the
boundary of $\Lambda$. The corresponding Gibbs state is defined by the expression
\eqref{defGS}, with $H_\Lambda$ replaced by $H_\Lambda + \sum_{A
\subset \Lambda} \partial_A^\Lambda$.

Ferromagnetic states are associated with operators of the
form $(n_{x\uparrow} - n_{x\downarrow})$ applied on the boundary of
the volume, while for antiferromagnetism we would use $(-1)^x (n_{x\uparrow} - n_{x\downarrow})$.
Of special importance are boundary conditions that break conservation
of the total number of particles. A state displaying superfluid
behaviour should be sensitive to the operator $\sum (\e{-\ii\theta}
c_x^\dagger + \e{\ii\theta} c_x)$ where the sum is over sites touching
the boundary. The order parameter for superfluidity is $c_0^\dagger$ (creation
operator of a particle at site 0) \cite{PO}, and with above boundary
conditions we may have $\expval{c_x^\dagger}^\theta = \alpha \e{\ii\theta}$
with $\alpha > 0$, revealing the presence of superfluidity.

For a superconductor with Cooper pairs described by a Hubbard-like
model, relevant boundary conditions are of the form $\sum
(\e{-\ii\theta} c_{x\uparrow}^\dagger c_{y\downarrow}^\dagger +
\e{\ii\theta} c_{x\uparrow} c_{y\downarrow})$ with the sum taken on
sites of the boundary, close to each other. This allows expectation
values of the form
$\expval{c_{0\uparrow}^\dagger c_{x\uparrow}^\dagger}$ to be non zero,
and this should be the indication of superconductivity \cite{Yang}.

The second way to obtain states with less symmetry than the
Hamiltonian is to add a perturbation, that is then set to zero. As for
superfluidity, a good perturbation to consider is $h \sum_{x \in
\Lambda} \bigl( \e{-\ii\theta} c_x^\dagger + \e{\ii\theta} c_x
\bigr)$ (see \eg \cite{Hua}), and the question is whether $\lim_{h \to 0}
\expval{c_0^\dagger}^{\theta,h}$ differs from zero.

Here we shall speak of {\it uniqueness of the Gibbs state} if it is
insensitive to both boundary conditions and to external
perturbations. In the range of parameters we consider, systems are
described by Gibbs states sharing the two properties
\begin{itemize}
\item $\lim_{\Lambda \nearrow \bbL} \expval
K_\Lambda^{\partial^\Lambda}$ does not depend on the boundary
conditions $\partial^\Lambda$, provided $\| \partial^\Lambda \|$ is small enough
(independently of $\Lambda$);
\item for all quantum perturbation $P$ with exponential decay, $\| P \|_c <
\infty$ for a large enough $c$, and all local observable $K$,
\be
\label{stability}
\expval K = \lim_{\alpha \to 0} \lim_{\Lambda \nearrow \bbL} \frac{\Tr K
\e{-\beta (H_\Lambda + \alpha \sum_{A \subset \Lambda} P_A)}}{\Tr \e{-\beta
(H_\Lambda + \alpha \sum_{A \subset \Lambda} P_A)}} .
\end{equation}
Notice that $P$ is not necessarily translation invariant, it may even not be
periodic.
\end{itemize}

Remark: the stability against perturbations can be given a simpler,
however more abstract definition. Let us consider $\caQ$, the Banach
space of interactions with finite norm \eqref{defnorm}, and $\caG$
the space of Gibbs states obtained with periodic boundary conditions;
$\caG$ is a topological space with the weak topology. Let $g$ denote
the corresponding mapping $\caQ \to \caG$. It is continuous at $H \in
\caQ$ provided $g^{-1}(G)$ is a neighbourhood of $H$ if $G$ is a
neighbourhood of $g(H)$.

Then the stability of a Gibbs state with respect to perturbations
amounts to saying that $g$ is continuous at $H$.

Indeed, we can see {\it ab absurdo} that \eqref{stability} implies
the continuity of $g$: suppose $G$ is a neighbourhood of $g(H)$ such
that $g^{-1}(G)$ is not a neighbourhood of $H$; since $\caQ$ is a
metric space, there exists a
sequence $(H_n)$, $H_n \to H$, with $H_n \notin g^{-1}(G)$; by \eqref{stability}, $g(H_n) \to g(H)$,
then $g(H_n) \in G$ for $n$ sufficiently large, and therefore $H_n \in
g^{-1}(G)$. Conversely, for any open set $G$ that contains $g(H)$,
$g^{-1}(G)$ is a neighbourhood of $H$; then if $H_n \to H$, we have
$H_n \in g^{-1}(G)$ for $n$ sufficiently large, therefore $g(H_n) \in G$.

\subsection{Result}

In order to state the result, we let $\beth$ be a constant that depends only on
the lattice, such that
\be
\#(A \ni x, \text{ connected}, |A| = k) \leq \beth^k .
\end{equation}
A possible choice, probably not optimal, is $\beth = (2\nu)^2$ for the
$\nu$-dimensional square lattice. The Golden Ratio appears here, that we write
$\phi = \frac{\sqrt5 + 1}2$ following a standard convention.\footnote{It is a
pleasure to welcome here the Golden Ratio. Its presence is however fortuitous
and does not involve any of its special and beautiful properties.}

\begin{theorem}[Analyticity in models with local interactions]\hfill
\label{thma}

Assume that $V^\bsmu$ and $T^\bsmu$ are smooth, \ie that all matrix elements of
$V^\bsmu_x$ and $T^\bsmu_A$ are analytic in $\bsmu$ for all $x$ and all $A$.
Let $c \geq c_0 = \log S + \log2\beth + \phi + 2\log\phi$. Then in the domain
$$
\beta \| T^\bsmu \|_c < 1 ,
$$
\begin{itemize}
\item[(i)] the free energy exists in the thermodynamic limit and is analytic in
$\beta$ and $\bsmu$;
\item[(ii)] the Gibbs state converges weakly in the thermodynamic limit;
\item[(iii)] the Gibbs state is exponentially clustering with a correlation
length bounded by $\xi = 4 (c-c_0)^{-1}$.
\end{itemize}
And the Gibbs state is unique, \ie
\begin{itemize}
\item[(iv)] the Gibbs state is stable with respect to boundary
conditions $\partial^\Lambda$ with $\beta \| T^\bsmu +
\partial^\Lambda \|_c < 1$ for all $\Lambda$;
\item[(v)] the Gibbs state is stable with respect to all external
perturbations $P$ with $\| P \|_c < \infty$.
\end{itemize}
\end{theorem}

Remark: the bound $4 (c-c_0)^{-1}$ for the correlation length is rather
arbitrary and could certainly be improved.

The stability against boundary conditions should hold for any bounded
$\partial^\Lambda$, not only small ones. However, such a statement is
difficult to prove in quantum systems, where we have to deal with
negative or complex numbers.

\begin{proof}[Proof of Theorem \ref{thma} (i)]

The idea of the proof is to expand the operator $\e{-\beta H_\Lambda}$ with
Duhamel formula; it allows next to express the partition
function as the one of a polymer model. After having shown that the weights of
polymers have exponential decay with respect to their size, the analyticity of
the free energy is a result of cluster expansions.

The Duhamel formula (very similar to the Trotter formula) yields
\bm
\Tr \e{-\beta H^\bsmu_\Lambda} = \Tr \e{-\beta \sum_{x \in \Lambda} V^\bsmu_x}
+ \sum_{m \geq 1} (-1)^m \sum_{A_1, \dots, A_m \subset \Lambda} \int_{0 <
\tau_1 < ... < \tau_m < \beta} \dd\tau_1 \dots \dd\tau_m \\
\Tr \e{-\tau_1 \sum_{x \in \Lambda} V^\bsmu_x} T_{A_1}^\bsmu \e{-(\tau_2 -
\tau_1) \sum_{x \in \Lambda} V^\bsmu_x} \dots T_{A_m}^\bsmu \e{-(\beta -
\tau_m) \sum_{x \in \Lambda} V^\bsmu_x} .
\end{multline}

For given $A_1, \dots, A_m$, we construct the graph $\caG$ of $m$ vertices,
with an edge between $i$ and $j$ whenever $A_i \cap A_j \neq \emptyset$.
Decomposing $\caG$ into connected subgraphs, it induces a partition of $\{ A_1,
\dots, A_m \}$ into $\ell$ subsets ($\ell \leq m$). We let $\caA_1, \dots,
\caA_\ell \subset \bbZ^\nu$ to be the unions of sets $A_1, \dots, A_m$ for each
partition. As a result, to each sequence $A_1, \dots, A_m$ corresponds a unique
set $\{ \caA_1, \dots, \caA_\ell \}$ of subsets of $\bbZ^\nu$, such that
\be
\begin{cases} \union_{i=1}^m A_i = \union_{i=1}^\ell \caA_i , & \\ \caA_i \cap
\caA_j = \emptyset & \text{if } i \neq j . \end{cases}
\end{equation}

We call $\caA_1, \dots,
\caA_\ell$ polymers and define their weight
\bm
\label{defrho}
\rho(\caA) = \e{\beta f_0(\beta, \bsmu) |\caA|} \sum_{m \geq 1} (-1)^m
\sum_{A_1, \dots, A_m} \sum_{n_\caA \in \Omega^\caA} \int_{0 < \tau_1 < ... <
\tau_m < \beta} \dd\tau_1 \dots \dd\tau_m \\
\bra{n_\caA} \e{-\tau_1 \sum_{x \in \caA} V^\bsmu_x} T_{A_1}^\bsmu
\e{-(\tau_2 - \tau_1) \sum_{x \in \caA} V^\bsmu_x} \dots T_{A_m}^\bsmu
\e{-(\beta - \tau_m) \sum_{x \in \caA} V^\bsmu_x} \ket{n_\caA} .
\end{multline}
The sum is over sets $A_1, \dots, A_m$ satisfying two restrictions: (i)
$\cup_{i=1}^m A_i = \caA$, (ii) the graph $\caG$ defined above is connected.
The partition function can then be rewritten as
\be
\label{partfct}
\Tr \e{-\beta H^\bsmu_\Lambda} = \e{-\beta f_0(\beta, \bsmu) |\Lambda|}
\sumtwo{\{ \caA_1, \dots, \caA_\ell \}}{\caA_i \cap \caA_j = \emptyset}
\prod_{j=1}^\ell \rho(\caA_j) .
\end{equation}

We have now to bound $\rho(\caA)$; first the matrix element:
\ba
\bigl| \bra{n_\caA} \cdot \ket{n_\caA} \bigr| &\leq \bigl\| \e{-\tau_1 \sum_{x
\in \caA} V^\bsmu_x} T_{A_1}^\bsmu \e{-(\tau_2 - \tau_1) \sum_{x \in \caA}
V^\bsmu_x} \dots T_{A_m}^\bsmu \e{-(\beta - \tau_m) \sum_{x \in \caA}
V^\bsmu_x} \bigr\| \nn\\
&\leq \bigl\| \e{-\beta V^\bsmu_x} \bigr\|^{|\caA|} \prod_{j=1}^m \|
T_{A_j}^\bsmu \| .
\end{align}
Let $e_0^\bsmu$ be the lowest eigenvalue of $V^\bsmu_x$; since $f_0(\beta,
\bsmu) \leq e_0^\bsmu$, we have
\be
\bigl\| \e{-\beta V^\bsmu_x} \bigr\| = \e{-\beta e_0^\bsmu} \leq \e{-\beta
f_0(\beta, \bsmu)} .
\end{equation}
Furthermore $|\Omega^\caA| = S^{|\caA|}$ and the integral over ``times"
$\{ \tau_j \}$ brings a factor $\beta^m / m!$; using $\| \caA \| \leq
\sum_{j=1}^m \| A_j \|$, we obtain
\ba
|\rho(\caA)| &\leq S^{|\caA|} \e{-c \| \caA \|} \sum_{m \geq 1}
\frac{\beta^m}{m!} \sum_{A_1, \dots, A_m \subset \caA} \prod_{j=1}^m \|
T_{A_j}^\bsmu \| \e{c \| A_j \|} \nn\\
&\leq S^{\| \caA \|} \e{-c \| \caA \|} \sum_{m \geq 1} \frac1{m!} \Bigl( \beta
|\caA| \sup_{x \in \bbZ^\nu} \sum_{A \ni x} \| T_A^\bsmu \| \e{c \| A \|}
\Bigr)^m \nn\\
&\leq \e{-(c - \log S - 1) \| \caA \|} .
\end{align}

Results of cluster expansions are summarized in Proposition \ref{propclexp}
below. From this we obtain the following expression for the free energy
\be
f(\beta, \bsmu) = f_0(\beta, \bsmu) - \frac1\beta \sum_{C, \supp C \ni x}
\frac{\Trfunc(C)}{|\supp C|} .
\end{equation}
It does not depend on $x$, because the Hamiltonian is translation invariant.
Since $\Trfunc(C)$ is analytic in $\beta$ and $\bsmu$, and the series converges
uniformly, the free energy $f(\beta, \bsmu)$ is an analytic function by Vitali
theorem.
\end{proof}

\begin{proposition}[Cluster expansions]\hfill
\label{propclexp}

Let us recall that we can choose $\beth = (2\nu)^2$ for the $\nu$-dimensional
square lattice and $\phi = \frac{\sqrt5 + 1}2$ is the Golden Ratio.

Assume that a function $z^{\beta, \bsmu} : \caP(\bbL) \to \bbC$ is given and
such that for all $A \subset \bbL$,
\begin{itemize}
\item $|z^{\beta, \bsmu}(A)| \leq \e{-\tau \| A \|}$ with $\tau \geq \tau_0 =
\log 2\beth + \phi-1 + 2\log\phi$;
\item $z^{\beta, \bsmu}(A)$ is analytic in $\beta, \bsmu$.
\end{itemize}

Then there exists an analytic function $\Trfunc : \caP(\caP(\bbL)) \to \bbC$
such that
$$
\log \sumtwo{\{ A_1, \dots, A_k \}}{A_j \subset \Lambda, A_i \cap A_j = \emptyset}
\prod_{j = 1}^k z^{\beta, \bsmu}(A_j) = \sumtwo{C = \{ A_1, \dots, A_k \}}{A_j
\subset \Lambda} \Trfunc(C) ;
$$
Let $\supp C = \cup_{A \in C} A$; $\Trfunc(C) = 0$ if $C$ is not a cluster, \ie
if $C = C_1 \cup C_2$ with $\supp C_1 \cap \supp C_2 = \emptyset$. This
function has exponential decay:
$$
\sumtwo{C = \{ A_1, \dots, A_k \}}{\cup_j A_j \ni x} |\Trfunc(C)| \e{(\tau -
\tau_0) \| C \|} \leq \phi-1
$$
for all $x \in \bbL$. Here, we set $\| C \| = \sum_{A \in C} \| A \|$.

\end{proposition}

This proposition is an immediate corollary of Koteck\'y and Preiss theorem on
cluster expansions \cite{KP} (see \cite{Dob} for an elegant and simpler proof).
To make the link between our notation and theirs:
\begin{table}[h]
\centering
\begin{tabular}{|c|c|}
\hline Here & \phantom{oo} \cite{KP} \phantom{oo} \\ \hline $A$ & $\gamma$ \\
$2\beth$ & $K$ \\ $(\phi-1) \| A \|$ & $a|\gamma|$ \\ $(\tau-\tau_0) \| A \|$ &
$d(\gamma)$ \\ \hline
\end{tabular}
\end{table}

Our polymers are not necessarily connected; but we note that their entropy
satisfies
\ba
\#(A \ni x, \| A \| = k) &\leq 2^k \, \#(A \ni x, \, \text{connected}, |A|=k)
\nn\\
&\leq (2\beth)^k . \nn
\end{align}
Hence the factor 2 in front of $\beth$.

A useful consequence of this proposition is the existence of the thermodynamic
limit of the free energy of a gas of polymers; if the weight $z^{\beta, \bsmu}$
is periodic with respect to lattice translations, then the limit
$$
f(\beta, \bsmu) = -\frac1\beta \lim_{\Lambda \nearrow \bbZ^\nu}
\frac1{|\Lambda|} \log \sumtwo{\{ A_1, \dots, A_k \}}{A_i \subset \Lambda, A_i
\cap A_j = \emptyset} \prod_{j = 1}^k z^{\beta, \bsmu}(A_j)
$$
exists and is analytic in $\beta$ and $\bsmu$.

\begin{proof}[Proof of Theorem \ref{thma} (ii)]

Having the expansion \eqref{partfct} for the partition function, the
treatment of expectation values of local operators is standard. The expectation value of $K$ involves the quantity $\Tr K
\e{-\beta H_\Lambda^\bsmu}$, that we expand as before with Duhamel formula.
\bm
\Tr K \e{-\beta H^\bsmu_\Lambda} = \Tr K \e{-\beta \sum_{x \in \Lambda}
V^\bsmu_x} + \sum_{m \geq 1} (-1)^m \sum_{A_1, \dots, A_m \subset \Lambda}
\int_{0 < \tau_1 < ... < \tau_m < \beta} \dd\tau_1 \dots \dd\tau_m \\
\Tr K \e{-\tau_1 \sum_{x \in \Lambda} V^\bsmu_x} T_{A_1}^\bsmu \e{-(\tau_2 -
\tau_1) \sum_{x \in \Lambda} V^\bsmu_x} \dots T_{A_m}^\bsmu \e{-(\beta -
\tau_m) \sum_{x \in \Lambda} V^\bsmu_x} .
\label{expK}
\end{multline}

We construct the graph $\caG$ of $(m+1)$ vertices, for the sets $\supp K, A_1,
\dots, A_m$, and we look for $(\ell+1)$ connected components. One of these
components contains $\supp K$, and we denote it by $\caA_K$; others are denoted
by $\caA_1, \dots, \caA_\ell$, as before. The weight of $\caA_K$ is modified by
the operator $K$; namely,
\bm
\label{defrhoK}
\rho_K(\caA_K) = \e{\beta f_0(\beta, \bsmu) |\caA_K|} \biggl\{ \Tr K \e{-\beta
\sum_{x \in \caA_K} V_x^\bsmu} + \sum_{m \geq 1} (-1)^m
\sum_{A_1, \dots, A_m} \sum_{n_{\caA_K} \in \Omega^{\caA_K}} \\
\int_{0 < \tau_1 <
... < \tau_m < \beta} \dd\tau_1 \dots \dd\tau_m
\bra{n_{\caA_K}} K \e{-\tau_1 \sum_{x \in \caA_K} V^\bsmu_x} T_{A_1}^\bsmu
\e{-(\tau_2 - \tau_1) \sum_{x \in \caA_K} V^\bsmu_x} \dots \\
\dots T_{A_m}^\bsmu
\e{-(\beta - \tau_m) \sum_{x \in \caA_K} V^\bsmu_x} \ket{n_{\caA_K}} \biggr\} .
\end{multline}
The expectation value of $K$ takes form
\be
\label{expansionK}
\expval K = \lim_{\Lambda \nearrow \bbL} \frac1{Z_\Lambda(\beta, \bsmu)}
\sum_{\{ \caA_K, \caA_1, \dots, \caA_\ell \}} \rho_K(\caA_K) \prod_{j=1}^\ell
\rho(\caA_j)
\end{equation}
where the sum is over non intersecting sets $\caA_K, \caA_1, \dots, \caA_\ell$,
$\ell \geq 0$. The weight $\rho_K$ has also exponential decay:
\be
|\rho_K(\caA_K)| \leq \| K \| \e{c \| \supp K \|} \e{-(c - \log S -1) \| \caA_K
\|} .
\end{equation}

From cluster expansions, we obtain an expression for $\expval K$:
\be
\expval K = \lim_{\Lambda \nearrow \bbL} \sumtwo{\caA_K \subset
\Lambda}{\caA_K \supset \supp K} \rho_K(\caA_K) \exp\Bigl\{ -\sumtwo{C, \supp C
\subset \Lambda}{\supp C \cap \caA_K \neq \emptyset} \Trfunc(C) \Bigr\} .
\end{equation}
The limit exists, because the sums converge uniformly in the volume $\Lambda$.
This proves the weak convergence of the Gibbs state in the thermodynamic limit.

\end{proof}

\begin{proof}[Proof of Theorem \ref{thma} (iii)]

Using cluster expansion, this is standard stuff. Expanding the expectation value $\expval{KK'}$ as before, we get
\bm
\label{defexpKK}
\expval{KK'} = \sum_{\caA_{KK'} \supset \supp K \cup \supp K'}
\rho_{KK'}(\caA_{KK'}) \exp\Bigl\{ -\sum_{C, \supp C \cap \caA_{KK'} \neq
\emptyset} \Trfunc(C) \Bigr\} \\
+ \sumthree{\caA_K \supset \supp K}{\caA_{K'} \supset \supp K'}{\caA_K \cap
\caA_{K'} = \emptyset} \rho_K(\caA_K) \rho_{K'}(\caA_{K'}) \exp\Bigl\{ -\sum_{C,
\supp C \cap [\caA_K \cup \caA_{K'}] \neq \emptyset} \Trfunc(C) \Bigr\} .
\end{multline}
The weights $\rho_K$, $\rho_{K'}$, $\rho_{KK'}$ are given by \eqref{defrhoK}
when substituting $K$ with $K, K', KK'$ respectively.

We define $\expval{KK'}^\short$ to be as \eqref{defexpKK}, but with sums only
over polymers and clusters of connected cardinality less than $\frac14 d(K,K')$
(we call such polymers {\it short}, they are {\it big} otherwise). We denote
$\expval{KK'}^\bbig = \expval{KK'} - \expval{KK'}^\short$.

When the expectation values are restricted to short polymers and clusters,
correlation functions are zero:
\be
\expval{KK'}^\short = \expval K^\short \expval{K'}^\short .
\end{equation}
Therefore
\be
\expval{KK'} - \expval K \expval{K'} = \expval{KK'}^\bbig - \expval K^\bbig
\expval{K'} - \expval K^\short \expval{K'}^\bbig .
\end{equation}
Expectation values $\expval\cdot^\bbig$ involve sums over polymers of
connected cardinality
bigger than $\frac14 d(K,K')$, and this has exponential decay. There are also
terms
\bm
\exp\Bigl\{ -\sum_{C, \supp C \cap \caA \neq \emptyset} \Trfunc(C) \Bigr\} -
\exp\Bigl\{ -\sumtwo{C, \supp C \cap \caA \neq \emptyset}{|\supp C| \leq
\frac14 d(K,K')} \Trfunc(C) \Bigr\} \\
= \exp\Bigl\{ -\sumtwo{C, \supp C \cap \caA \neq \emptyset}{|\supp C| \leq
\frac14 d(K,K')} \Trfunc(C) \Bigr\} \Bigl[ \exp\Bigl\{ -\sumtwo{C, \supp C \cap
\caA \neq \emptyset}{|\supp C| > \frac14 d(K,K')} \Trfunc(C) \Bigr\} - 1 \Bigr]
.
\end{multline}
We know from Proposition \ref{propclexp} that the sum over clusters has
exponential decay; more precisely, the quantity between brackets is bounded by
$C(\caA) \e{-\frac14 d(K,K') (c-c_0)}$, where $C(\caA)$ depends on $|\caA|$
only.

Exponential clustering is now clear.

\end{proof}

\begin{proof}[Proof of Theorem \ref{thma} (iv)]

The proof is rather standard, so we content ourselves by outlining
it.

Expanding $\Tr \e{-\beta H_\Lambda^\bsmu - \beta \sum_{A \subset
\Lambda} \partial_A^\Lambda}$ with Duhamel formula, we obtain an
expression very similar to \eqref{expK}. The difference is that
operators $\partial_A^\Lambda$ now appear in the second line of
\eqref{expK}. We define $\hat\rho_K(\caA_K)$ to be as in
\eqref{defrhoK}, except that at least one boundary operator
$\partial_A^\Lambda$ shows up. Similarly, let $\hat\rho(\caA)$ be like
\eqref{defrho} but with at least one $\partial_A^\Lambda$; an
important property of these weights is that they are zero if the
polymer does not touch the boundary of $\Lambda$. As a
result we get
\be
\expval K_\Lambda^{\partial^\Lambda} = \sumtwo{\caA_K \subset
\Lambda}{\caA_K \supset \supp K} \Bigl( \rho_K(\caA_K) +
\hat\rho_K(\caA_K) \Bigl) \exp\Bigl\{ -\sumtwo{C, \supp C \subset
\Lambda}{\supp C \cap \caA_K \neq \emptyset} \bigl( \Trfunc(C) +
\hat\Trfunc(C) \bigr) \Bigr\}
\end{equation}
where $\hat\Trfunc(C)$ is such that $[\Trfunc(C) + \hat\Trfunc(C)]$ is
the truncated function for polymers with weights $[\rho(\caA) +
\hat\rho(\caA)]$.

A few more developments lead to an expression for $\expval
K_\Lambda^{\partial^\Lambda}$ that is equal to $\expval
K_\Lambda$, plus terms that connect $\supp K$ with the boundary of
$\Lambda$, and that decay exponentially quickly. In the thermodynamic limit, this
correction vanishes, and therefore for all $\partial^\Lambda$:
\be
\lim_{\Lambda \nearrow \bbL} \expval K_\Lambda^{\partial^\Lambda} =
\expval K .
\end{equation}

\end{proof}

\begin{proof}[Proof of Theorem \ref{thma} (v)]

Once we have an expansion in terms of clusters, the proof of stability
with respect to external perturbations is actually fairly simple.

First we note that if $\sum_{n \geq 0} |g_n^\alpha| < \infty$ uniformly in
$\alpha$, and $g_n^\alpha \to g_n$ when $\alpha \to 0$, then
$$
\sum_{n \geq 0} g_n^\alpha \to \sum_{n \geq 0} g_n .
$$
Indeed, for any $\varepsilon > 0$ there exists $m$ such that
$$
\sum_{n \geq m} \bigl( |g_n^\alpha| + |g_n| \bigr) \leq \frac\varepsilon2 .
$$
Furthermore, for all $n$ there exists $\bar\alpha_n > 0$ such that $|g_n^\alpha
- g_n| \leq \frac\varepsilon{2m}$ when $\alpha \leq \bar\alpha_n$. Choosing
$\bar\alpha = \min_{n < m} \bar\alpha_n$, we have for all $\alpha \leq
\bar\alpha$
$$
\Bigl| \sum_{n \geq 0} g_n^\alpha - \sum_{n \geq 0} g_n \Bigr| \leq \sum_{n<m}
|g_n^\alpha - g_n| + \sum_{n \geq m} \bigl( |g_n^\alpha| + |g_n| \bigr) \leq
\varepsilon .
$$

We add to the Hamiltonian a new quantum interaction $P$ with $\| P \|_c <
\infty$. Since $\| T^\bsmu \|_c < 1$, there exists $\bar\alpha > 0$ such that
for all $\alpha \leq \bar\alpha$,
\be
\| T^\bsmu + \alpha P \|_c < 1 .
\end{equation}
Retracing the steps above, we obtain
\be
\label{defKalpha}
\expval K^\alpha = \sum_{\caA_K \supset \supp K} \rho_K^\alpha(\caA_K)
\exp\Bigl\{ -\sum_{C, \supp C \cap \caA_K \neq \emptyset} \Trfunc_\alpha(C)
\Bigr\}
\end{equation}
where $\rho_K^\alpha(\caA_K)$ is obtained by replacing $T^\bsmu_A$ with
$(T^\bsmu_A + \alpha P_A)$ in \eqref{defrhoK}; similarly, $\Trfunc_\alpha$ is
constructed with weights $\rho^\alpha$ that we get by the same substitution in
\eqref{defrho}.

The weights $\rho^\alpha$ and $\rho_K^\alpha$ are absolutely convergent series in matrix
elements of $\{ T^\bsmu_A + \alpha P_A \}$, therefore $\rho^\alpha \to \rho$
and $\rho_K^\alpha \to \rho_K$ as $\alpha \to 0$. Hence $\Trfunc_\alpha \to
\Trfunc$, and also
\be
\sum_{C, \supp C \cap \caA_K \neq \emptyset} \Trfunc_\alpha(C) \to \sum_{C,
\supp C \cap \caA_K \neq \emptyset} \Trfunc(C) .
\end{equation}
Since the expression \eqref{defKalpha} for $\expval K^\alpha$ is absolutely
convergent, this implies that $\expval K^\alpha \to \expval K$.

\end{proof}

\section{The Hubbard model}

The phase space of the Hubbard model is the Fock space of antisymmetric wave
functions on $\Lambda \subset \bbL$. A convenient basis is the one in
occupation numbers of position operators. It is given by $\{ \ket{n_\Lambda}
\}_{n_\Lambda \in \Omega^\Lambda}$ where $\Omega = \{ 0, \uparrow, \downarrow, 2
\}$. The Hamiltonian is
\be
H_\Lambda = -t \sumtwo{\neighbours xy \subset \Lambda}{\sigma \in \{ \uparrow,
\downarrow \}} c^\dagger_{x\sigma} c_{y\sigma} + U \sum_{x \in \Lambda}
n_{x\uparrow} n_{x\downarrow} - \mu \sum_{x \in \Lambda} (n_{x\uparrow} +
n_{x\downarrow})
\end{equation}
where the first sum is over nearest neighbours $x, y \in \Lambda$. The first
term represents the kinetic energy, the second one is the local repulsion
between electrons, and the last term is the chemical potential multiplying the
total number of electrons.

\bfig
$$
\begin{matrix}
\hspace{5mm} \epsfxsize=30mm \epsffile{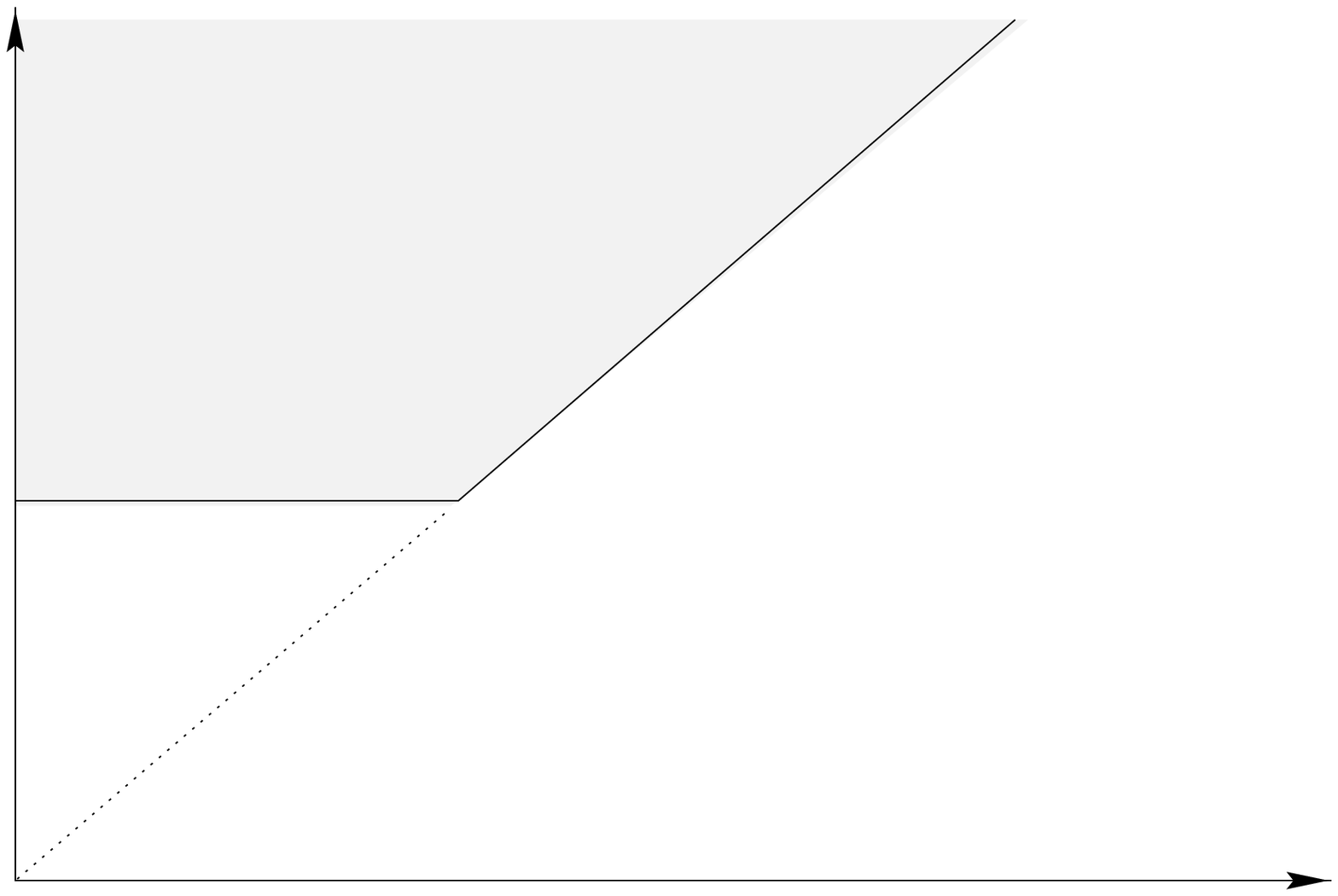} \hspace{5mm} & \hspace{5mm}
\epsfxsize=30mm \epsffile{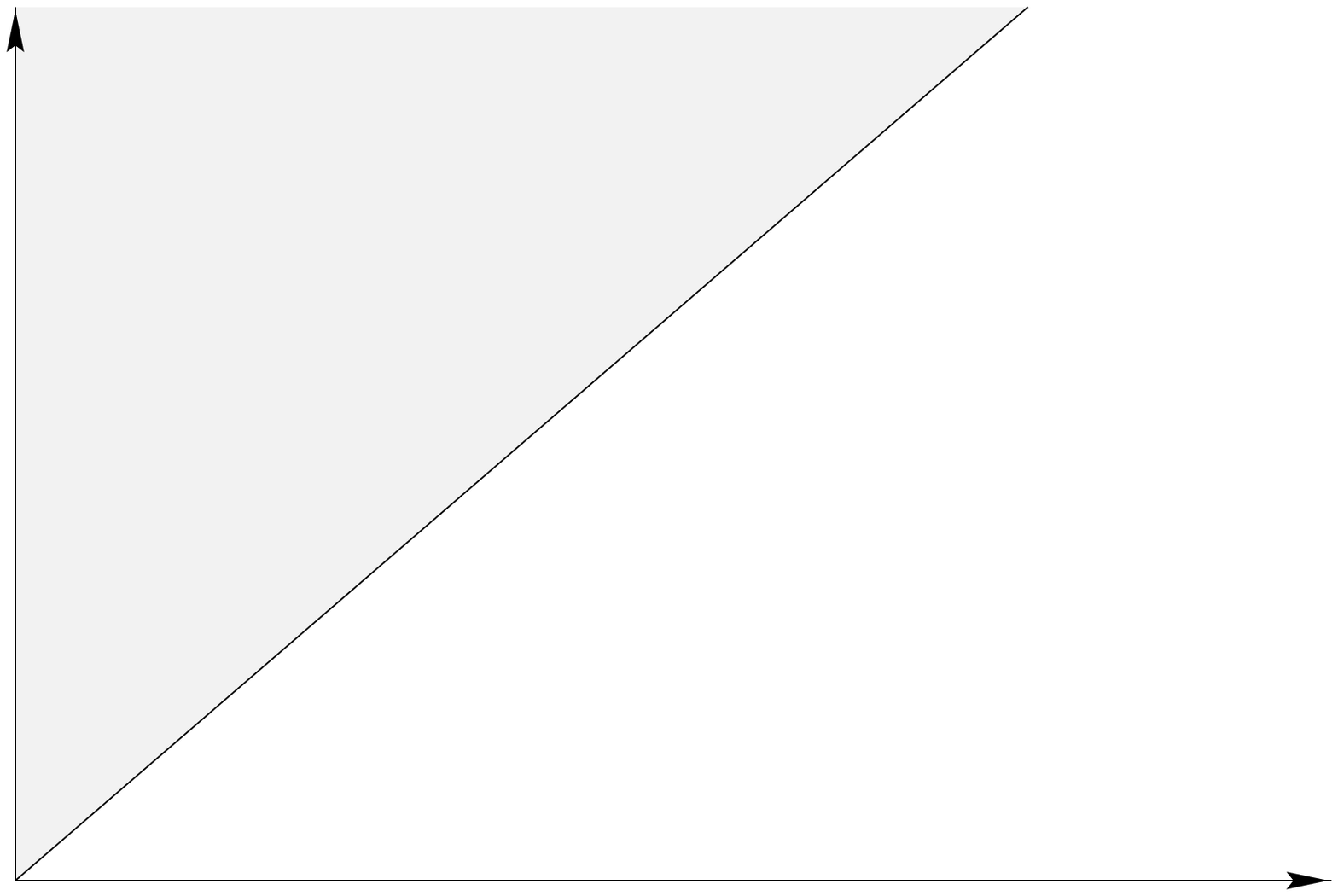} \hspace{5mm} & \hspace{5mm} \epsfxsize=30mm
\epsffile{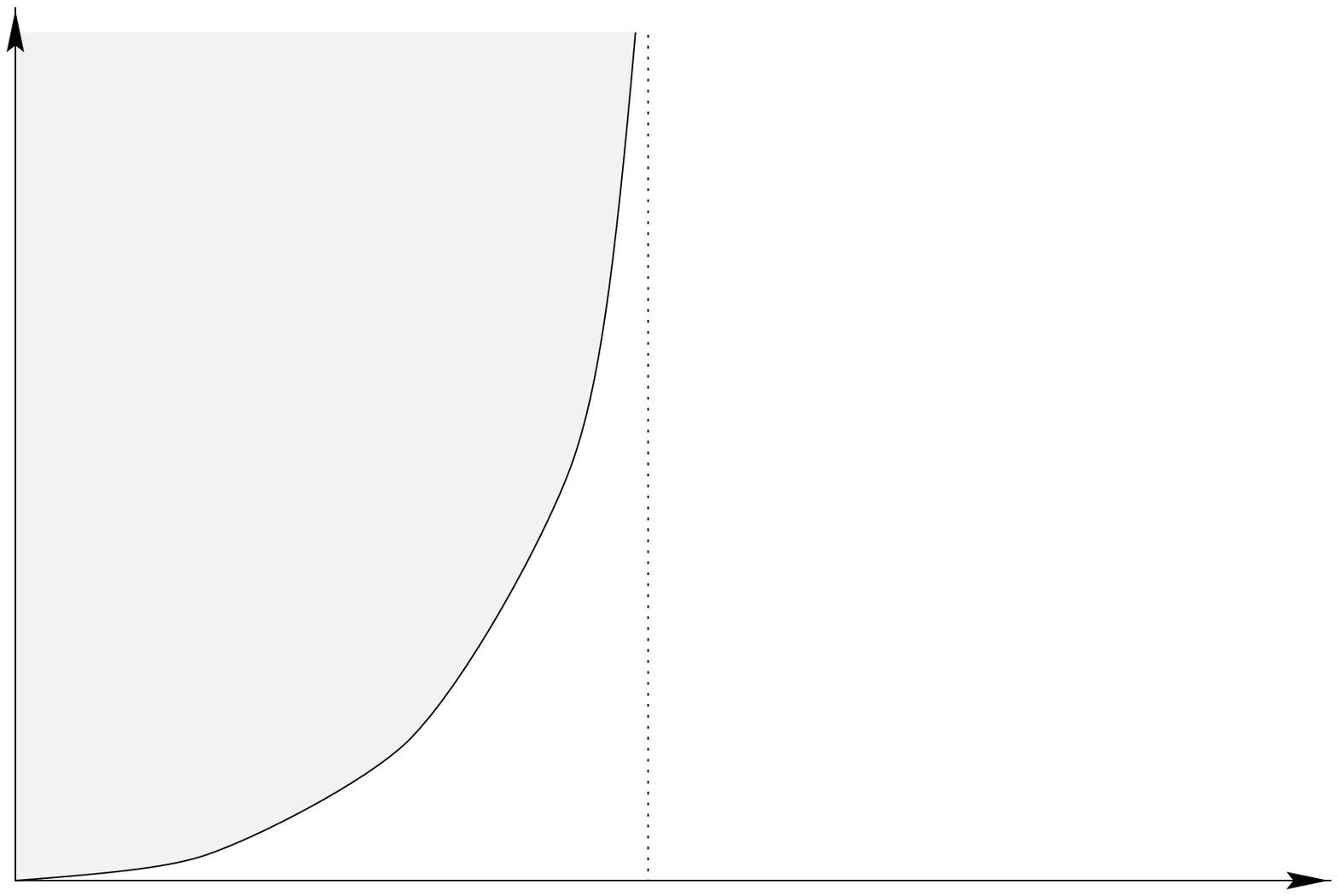} \hspace{5mm} \\
\text{\footnotesize (a)} & \text{\footnotesize (b)} & \text{\footnotesize (c)}
\end{matrix}
$$
\figtext{
\writefig   1.2  2.5  {\tiny $\frac1\beta$}
\writefig   4.6  0.65 {\tiny $t$}
\writefig   5.6  2.5  {\tiny $\frac1\beta$}
\writefig   8.9  0.65 {\tiny $t$}
\writefig   9.9  2.5  {\tiny $\frac1\beta$}
\writefig  13.3  0.65 {\tiny $t$}
}
\caption{Domains of analyticity, stemming from (a) high temperature expansions,
(b) domain $\caD_1$ of Theorem \ref{thmHubbard}, and (c) domain $\caD_2$ of the
same theorem.}
\label{threephd}
\end{figure}

High temperature expansions yield analyticity of the free energy for all $\beta$
such that
$$
\beta t < \const \hspace{3mm} \text{ and } \hspace{3mm} \beta U < \const ,
$$
see \fig 1; the domain of analyticity may be extended, as we see now. Let $\chi$ be the
maximum coordination number of $\bbL$ ($\chi = 2\nu$ for $\nu$-dimensional
square lattice), and $\epsilon = (2\chi\beth \e\phi \phi^2)^{-1}$.

\begin{theorem}[Analyticity in the Hubbard model]\hfill
\label{thmHubbard}

Let $\Delta = \min(\mu, U-\mu)$. Thermodynamic limits of the free energy and of
the Gibbs state exist in the domain $\caD_1 \cup \caD_2$, where
$$
\caD_1 = \bigl\{ (\beta, t, \mu) : \beta t < \epsilon \bigr\}
$$
and
$$
\caD_2 = \bigl\{ (\beta, t, \mu) : 0<\mu<U \text{ and } \beta \tfrac{t^2}\Delta
< 2\chi \epsilon^2 (1 - 2t / \epsilon \Delta) \bigr\} .
$$

The free energy is analytic, and the Gibbs state is exponentially clustering and
unique (that is, stable against boundary conditions
$\partial^\Lambda$, $\| \partial^\Lambda \|_c < 1$, and external
perturbations $P$, $\| P \|_c < \infty$ for a sufficiently big $c$).
\end{theorem}

Remark that the domain $\caD_2$ is meaningful only if $t$ is small enough,
namely $\tfrac t\Delta < \tfrac12 \epsilon$. For the 3-dimensional square
lattice, we find for $\caD_1$ the condition $\beta t < 1.75... \cdot 10^{-4}$,
and for $\caD_2$, $\frac{\beta t^2}\Delta < 3.68... \cdot 10^{-7} (1 - 1.14...
\cdot 10^4 \frac t\Delta)$. Of course, the domain of analyticity is much larger
than these domains, where analyticity is proven to hold.

These properties likely hold for all $\beta <
\infty$ in dimension 1, and possibly also in dimension 2. When $\nu \geq 3$, a domain with antiferromagnetic
phase is expected for $t \ll U$ and $\beta t^2 /U > \const$. Such a phase can be
proven in the {\it asymmetric} Hubbard model, where electrons of different
spins are assumed to have different hopping parameters \cite{KL, LM, MM,
DFF} (see also \cite{DFFR} and
\cite{KU} for two general methods to study rigorously such situations). Assuming
this to be true in the standard Hubbard model, we observe that the
condition for the domain
$\caD_2$ is qualitatively correct; this is illustrated in \fig
\ref{phdHubbard}.

\bfig
\epsfxsize=60mm
\centerline{\epsffile{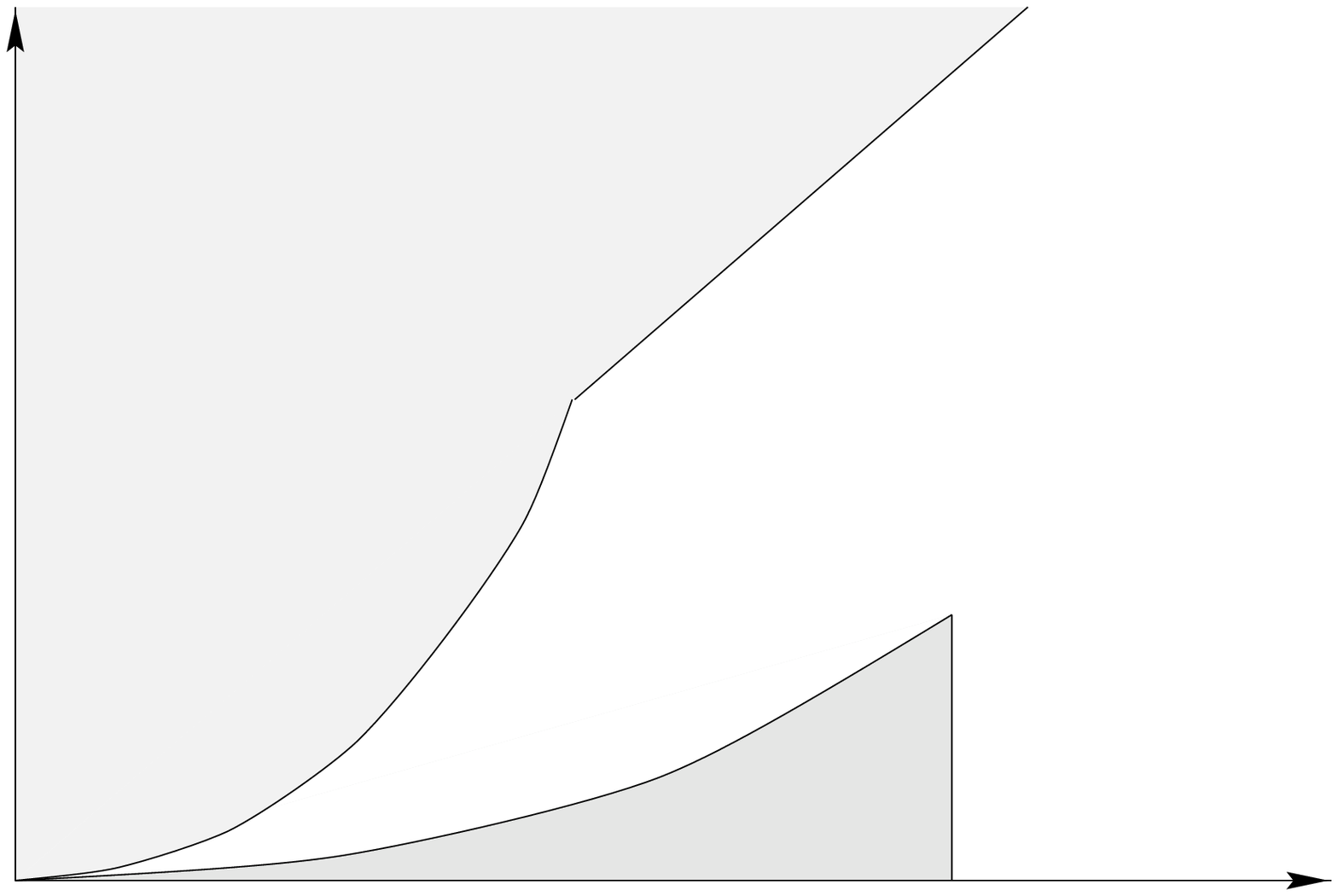}}
\figtext{
\writefig   2.7  4.2  {\tiny temperature}
\writefig  10.45 0.35 {$t$}
\writefig   4.8  3.5  {\it\footnotesize uniqueness}
\writefig   6.38 0.6  {\it\tiny antiferromagnetic}
}
\caption{Phase diagram of the Hubbard model. Antiferromagnetic phase is
expected for dimension $\nu \geq 3$; it can be proven when $\nu \geq 2$ for the
{\it asymmetric} model.}
\label{phdHubbard}
\end{figure}

\begin{proof}[Proof of Theorem \ref{thmHubbard}, domain $\caD_1$]

The classical free energy of the Hubbard model is easily computed and
is given by
\be
f_0(\beta, \mu) = -\frac1\beta \log \bigl[ 1 + 2\e{\beta\mu} + \e{-\beta U +
2\beta\mu} \bigr] .
\end{equation}

Let us write the kinetic operator $-t \sum_{\bsA \subset \Lambda} T_\bsA$ where
$\bsA = (\neighbours xy, \sigma)$ and the notation $\bsA \subset \Lambda$ means
$x,y \in \Lambda$; $T_\bsA = c^\dagger_{x\sigma} c_{y\sigma}$.

The expression \eqref{defrho} for $\rho(\caA)$ takes the following form
\bm
\rho(\caA) = \e{\beta f_0(\beta, \mu) |\caA|} \sum_{m \geq 1} t^m \sum_{\bsA_1,
\dots, \bsA_m} \sum_{n_\caA \in \Omega^\caA} \int_{0 < \tau_1 < ... < \tau_m <
\beta} \dd\tau_1 \dots \dd\tau_m \\
\bra{n_\caA} \e{-\tau_1 \sum_{x \in \caA} V^\mu_x} T_{\bsA_1} \e{-(\tau_2 -
\tau_1) \sum_{x \in \caA} V^\mu_x} \dots T_{\bsA_m} \e{-(\beta - \tau_m)
\sum_{x \in \caA} V^\mu_x} \ket{n_\caA}
\end{multline}
with a restriction on the sum over $\bsA_1, \dots, \bsA_m$, namely their union
yields $\caA$ and they are connected in the sense of the graph $\caG$ described
above. Notice that here $\rho(\caA) = 0$ if $\caA$ is not connected. The
expression for $\rho_K(\caA_K)$ is similar, compare with \eqref{defrhoK}.

We obtain the domain $\caD_1$ of Theorem \ref{thmHubbard} by proceeding as
before. Namely, we bound the matrix element with $\e{-\beta e_0^\mu |\caA|}
\leq \e{-\beta f_0(\beta, \mu) |\caA|}$. A few observations allow to slightly
optimize the bound for $\rho(\caA)$. First, there are no more than $\frac\chi2
|\caA|$ sets of nearest neighbours in $\caA$. Second, if we choose $\bsA_1,
\dots, \bsA_m$ (such that they cover $\caA$), then there is at most one
configuration $n_\caA$ such that $\bra{n_\caA} T_{\bsA_1} \dots T_{\bsA_m}
\ket{n_\caA}$ differs from 0.

Therefore we obtain the bound
\ba
|\rho(\caA)| &\leq \e{-(c+1) |\caA|} \sum_{m \geq 0} \frac{(\beta t)^m}{m!}
(\frac\chi2 |\caA|)^m 4^m \e{(c+1) m} \nn\\
&\leq \e{-c |\caA|} ,
\end{align}
assuming that
\be
\label{condition}
2\chi \e{c+1} \beta t \leq 1 .
\end{equation}
Here the polymers are connected sets; in this case, Proposition \ref{propclexp}
holds with $2\beth$ replaced by $\beth$. Therefore the condition
\eqref{condition} must be fulfilled with $c_0 = \log\beth + \phi -1 +
2\log\phi$. When this inequality is strict, it also holds with $c > c_0$, so
that we obtain exponential clustering and stability against
perturbations or boundary interactions.

\end{proof}

\begin{proof}[Proof of Theorem \ref{thmHubbard}, domain $\caD_2$]

Domain $\caD_2$ benefits from the following geometric representation (see
\fig\ref{figloops} for intuition). First we let $n^m_\caA = n_\caA$, then
$\ket{n^{m-1}_\caA} = \pm T_{\bsA_m} \ket{n^m_\caA}$, \dots, $\ket{n^1_\caA} =
\pm T_{\bsA_2} \ket{n^2_\caA}$, $\ket{n^m_\caA} = \pm T_{\bsA_1}
\ket{n^1_\caA}$. The last condition follows by cyclicity of the trace.

\bfig
\epsfxsize=80mm
\centerline{\epsffile{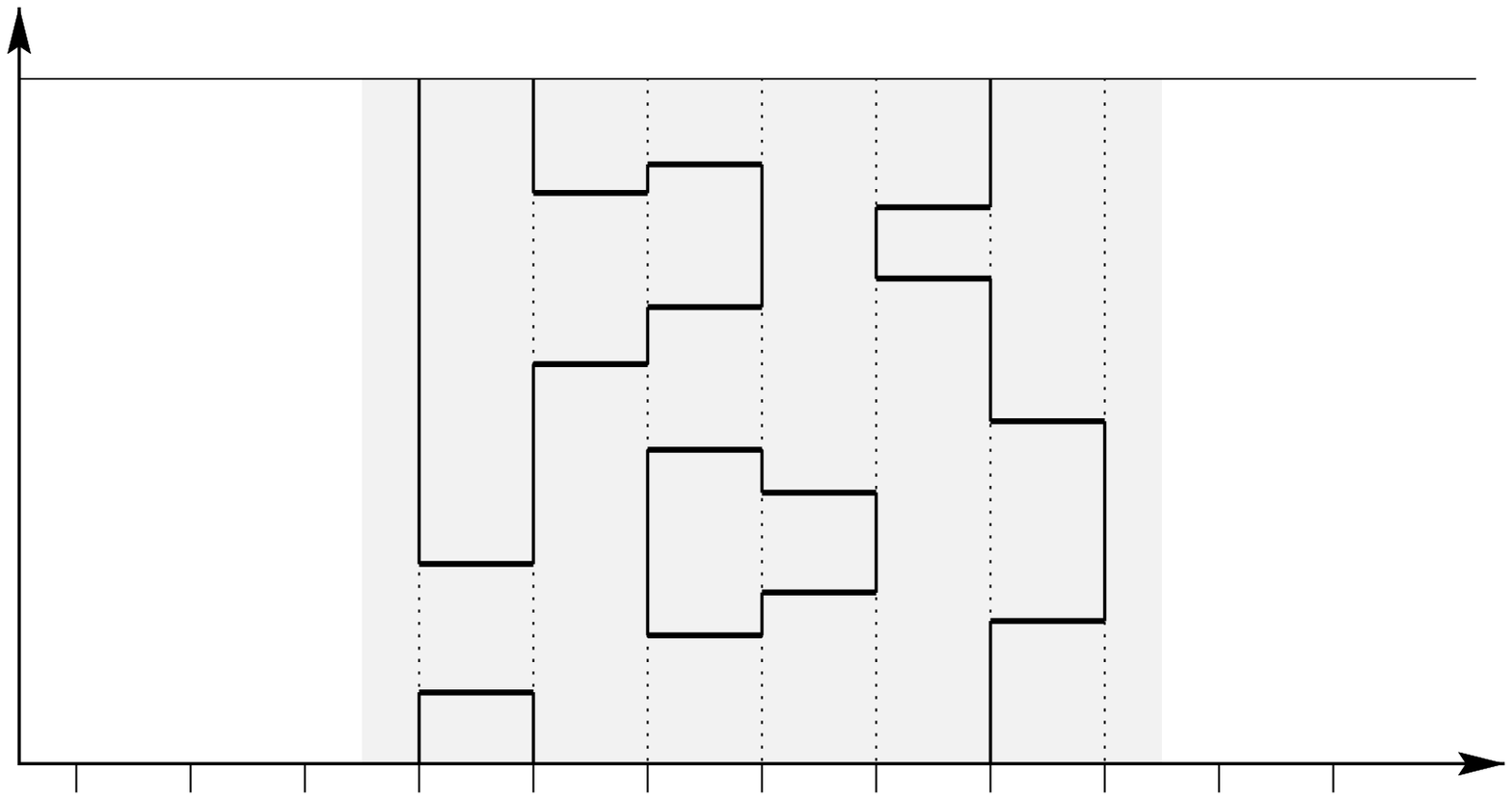}}
\figtext{
\writefig  11.24 0.2  {$\caA \subset \Lambda^\nu$}
\writefig   2.99 4.2  {$\beta$}
\writefig   5.47 0.2  {\footnotesize $0$}
\writefig   6.08 0.2  {\footnotesize $2$}
\writefig   6.69 0.2  {\footnotesize $\uparrow$}
\writefig   7.31 0.2  {\footnotesize $\downarrow$}
\writefig   7.91 0.2  {\footnotesize $\downarrow$}
\writefig   8.51 0.2  {\footnotesize $2$}
\writefig   9.12 0.2  {\footnotesize $\uparrow$}
\writefig   5.42 1.3  {\tiny $\uparrow$}
\writefig   6.03 1.3  {\tiny $\downarrow$}
\writefig   6.03 2.1  {\tiny $2$}
\writefig   6.03 3.2  {\tiny $\downarrow$}
\writefig   6.63 1.6  {\tiny $2$}
\writefig   6.63 2.5  {\tiny $\downarrow$}
\writefig   6.63 2.82 {\tiny $2$}
\writefig   6.63 3.3  {\tiny $\uparrow$}
\writefig   6.63 3.71 {\tiny $0$}
\writefig   7.25 1.36 {\tiny $0$}
\writefig   7.25 1.7  {\tiny $\downarrow$}
\writefig   7.25 2.13 {\tiny $0$}
\writefig   7.25 2.5  {\tiny $\uparrow$}
\writefig   7.25 3.3  {\tiny $2$}
\writefig   7.85 1.7  {\tiny $0$}
\writefig   7.85 2.4  {\tiny $\downarrow$}
\writefig   7.85 3.35 {\tiny $2$}
\writefig   8.45 1.8  {\tiny $\uparrow$}
\writefig   8.45 2.8  {\tiny $2$}
\writefig   8.45 3.35 {\tiny $\downarrow$}
\writefig   9.07 1.8  {\tiny $2$}
}
\caption{Three loops.}
\label{figloops}
\end{figure}

Next, if $\bsA_j = (\neighbours{x_j}{y_j}, \sigma)$, we define horizontal bonds
$\caB \subset \bbR^\nu \times [0, \beta]_\per$
$$
\caB = \union_{j=1}^m \overline{x_j y_j} \times \{ \tau_j \}
$$
where $\overline{x_j y_j} \subset \bbR^\nu$ is the segment joining $x_j$ and
$y_j$. We consider vertical segments
$$
\caS = \union_{j=0}^m \bigl\{ x \in \bbL: n^j_x \in \{ 0,2 \} \bigr\} \times
[\tau_j, \tau_{j+1}] ,
$$
where we set $\tau_0=0$, $\tau_{m+1}=\beta$ and $n^0=n^m$. Actually, $\caS$ is
a subset of $\bbL \times [0, \beta]_\per$; but with a small abuse of notation,
we consider $\caS \subset \bbR^\nu \times [0, \beta]_\per$.

The set $\caB \cup \caS$ decomposes into a finite number of closed circuits
that we call {\it loops}.\footnote{This representation has many similarities
with that of \cite{MM}, introduced for the Falicov-Kimball model.} To be
precise, a loop $\ell$ is a pair $(\supp\ell, \bsA(\ell))$ where $\supp\ell
\subset \bbR^\nu \times [0, \beta]_\per$ is the support of $\ell$, and
$\bsA(\ell) = (\bsA_1, \dots, \bsA_{m(\ell)})$ are successive applications of
operators $T_{\bsA_1}, \dots, T_{\bsA_{m(\ell)}}$; here $m(\ell)$ is the number
of horizontal segments (``jumps") in $\ell$ that we mark out $1, \dots, m$ in
increasing vertical coordinates, and $T_{\bsA_j}$ is the operator associated
with the segment $j$.

The weight $\rho(\caA)$ can be written as an integral over sets of loops, with
many restrictions. In particular, each vertical line $\{ x \} \times [0,
\beta]_\per$, $x \in \caA$, must intersect at least one loop. As a consequence,
to a given set of loops corresponds at most one sequence of configurations
$(n^1_\caA, \dots, n^m_\caA)$.
\be
\rho(\caA) = \e{\beta f_0(\beta, \mu) |\caA|} \e{\beta\mu |\caA|} \sum_{k \geq
1} \frac1{k!} \int \dd\ell_1 \dots \dd\ell_k \, \varepsilon(\ell_1, \dots, \ell_k)
\prod_{j=1}^k z(\ell_j)
\end{equation}
where
\be
\varepsilon(\ell_1, \dots, \ell_k) = \bra{n_\caA} \prod_{\bsA \in (\ell_1,
\dots, \ell_k)} T_\bsA \ket{n_\caA}
\end{equation}
and
\be
z(\ell) = t^{m(\ell)} \e{-\mu |\ell|_0 - (U-\mu) |\ell|_2} .
\end{equation}
Let us explain these notations. The configuration $n_\caA$ is defined by $(\ell_1,
\dots, \ell_k)$; namely, if $\{ x \} \times \{ 0 \} \in \supp\ell_j$, we know
that $n_x \in \{ 0,2 \}$; looking at the first occurrence of an operator
$T_\bsA$, $\bsA \ni x$, we can check whether a particle is created or
annihilated at $x$, in which case $n_x = 0$ or $n_x = 2$ respectively.
Similarly, if $\{ x \} \times \{ 0 \} \notin \cup_j\supp\ell_j$, we have $n_x
\in \{ \uparrow, \downarrow \}$; if the first operator $T_\bsA$ such that $\bsA
\ni x$ creates an $\uparrow$ electron, or annihilate a $\downarrow$ electron, we
have $n_x = \downarrow$; otherwise $n_x = \uparrow$.

The product is over all operators $T_\bsA$ that occur in the loops, ordered in
decreasing vertical coordinate of the corresponding horizontal segment. Notice
that $\varepsilon(\cdot) \in \{ -1, 0, 1 \}$.

The vertical length of a loop is $|\ell| = |\ell|_0 + |\ell|_2$, where
$|\ell|_j$ denotes the length of all vertical segments where the configuration
takes value $j$. We bound
\be
z(\ell) \leq t^{m(\ell)} \e{-\Delta |\ell|} .
\end{equation}

We have the following bound for $\rho(\caA)$ (we use $f_0(\beta, \mu) \leq
\mu$):
\be
|\rho(\caA)| \leq \e{-(c+1) |\caA|} \sum_{k \geq 1} \frac1{k!} \Bigl[
\int_{\supp\ell \subset \caA \times [0, \beta]_\per} \dd\ell (t
\e{c+1})^{m(\ell)} \e{-\Delta |\ell|} \Bigr]^k .
\end{equation}
The integral over one loop with $m$ jumps may be evaluated in the following way.
\begin{enumerate}
\item We choose two nearest neighbour sites in $\caA$ (there are less than $\chi
|\caA|$ possibilities), we integrate over a number $\tau$ in $[0,\beta]$, and we
choose a spin; we obtain the first jump of the loop.
\item We decide whether the loop is going up or down in the vertical dimension,
we integrate over the vertical distance, we choose a neighbour of our site and a
spin; integration over the vertical distance is bounded by
$$
\int_0^\infty \dd\tau \e{-\Delta \tau} = \frac1\Delta .
$$
We repeat this procedure until the last jump but one.
\item For the last jump, we decide whether the vertical direction is up or down,
and we integrate over the distance, yielding a factor $1/\Delta$. Then the loop
completes itself in a unique way (provided there is a way); therefore there are
no sums over nearest neighbour and spin.
\end{enumerate}
Since the first jump is arbitrary, we can divide by $m$ the contribution of
loops with $m$ jumps. Notice that the second step is superfluous when $m=2$. We
obtain
\ba
\sum_{m \geq 2} \inttwo{\supp\ell \subset \caA \times [0,\beta]_\per}{m(\ell) =
m} \dd\ell (t \e{c+1})^m
\e{-\Delta |\ell|} &\leq \sum_{m \geq 2} \frac1m (t \e{c+1})^m \chi |\caA| \beta
2 [2 \frac1\Delta \chi 2]^{m-2} 2 \frac1\Delta \nn\\
&\leq |\caA| \frac{\beta t^2}\Delta 2\chi \e{2(c+1)} \frac1{1 - 4\chi \e{c+1}
\frac t\Delta} .
\end{align}
From the conditions $\frac t\Delta < \frac1{4\chi} \e{-(c+1)}$ and $\frac{\beta
t^2}\Delta < \frac1{2\chi} \e{-2(c+1)} (1 - 4\chi \e{c+1} \frac t\Delta)$, we
finally have
\be
|\rho(\caA)| \leq \e{-(c+1) |\caA|} \sum_{k \geq 1} \frac1{k!} |\caA|^k \leq
\e{-c|\caA|}
\end{equation}
with $c = \log\beth + \phi-1 + 2\log\phi$. Since the weights of polymers are
analytic functions of $\beta, \mu$, so is the free energy in the thermodynamic
limit.

Existence and properties of the Gibbs state are readily obtained by repeating
the proofs of Section 2, using above estimates.

\end{proof}

\section{The Bose-Hubbard model}

The Bose-Hubbard model describes a lattice system of interacting bosons. In a
finite volume $\Lambda \in \bbL$, the phase space is the Hilbert space with
basis $\{ \ket{n_\Lambda} : n_\Lambda \in \bbN^\Lambda \}$; the Hamiltonian
consists in a kinetic operator and a local repulsive interaction:
\be
H_\Lambda = t \sum_{\neighbours xy \subset \Lambda} c^\dagger_x c_y + U \sum_{x
\in \Lambda} (\hat n_x^2 - \hat n_x) -\mu \sum_{x \in \Lambda} \hat n_x .
\end{equation}
The first term is a standard hopping operator between nearest-neighbours; the
second term describes the local repulsion between bosons (each pair of particles
at a given site contributes for $2U$); the chemical potential $\mu \in \bbR$
controls the density of the system.

\begin{figure}[h]
\epsfxsize=60mm
\centerline{\epsffile{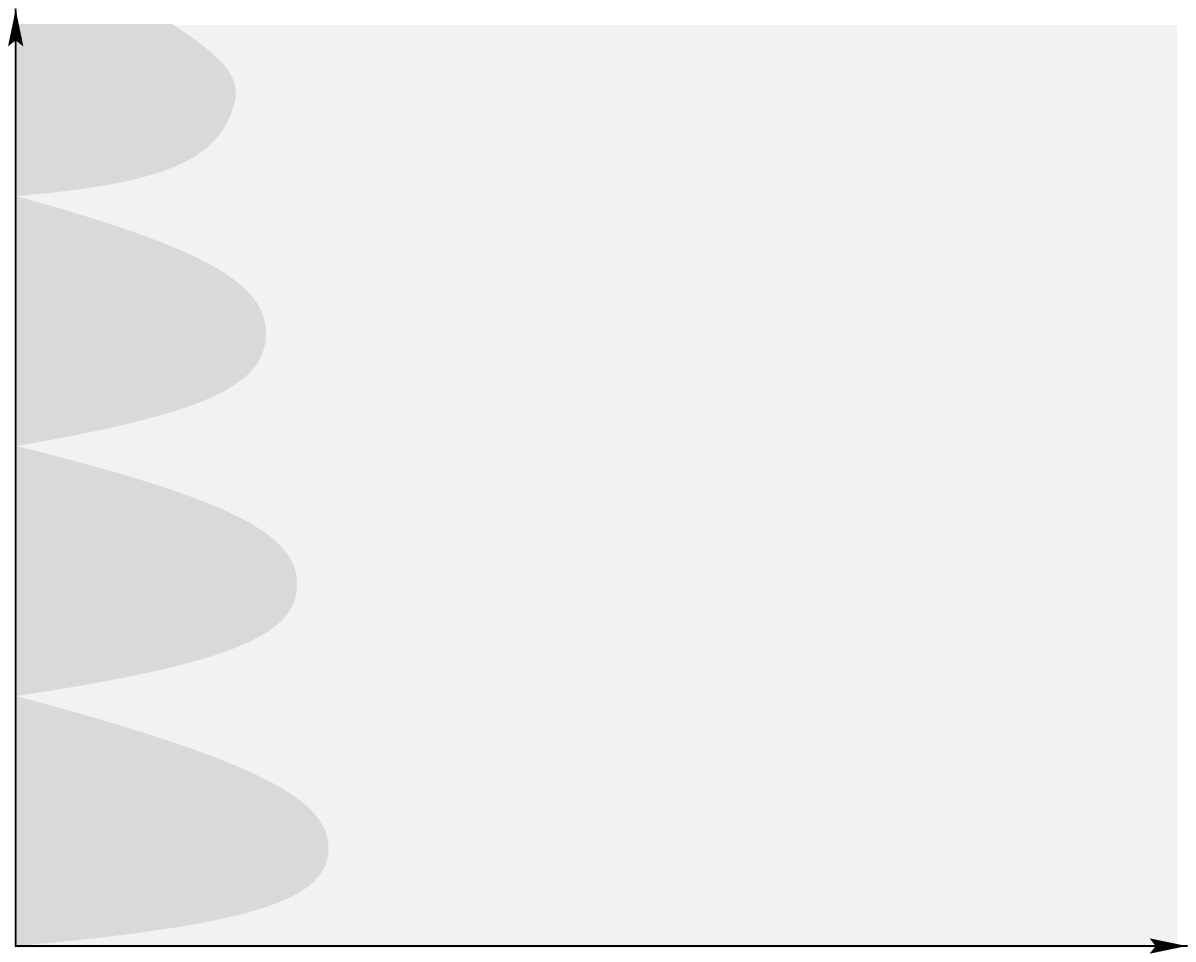}}
\figtext{
        \writefig      10.45	0.4	{$t$}
        \writefig       4.0	5.1	{$\mu$}
 	\writefig       7.0	3.2	{\it superfluid}
        \writefig	4.65	1.0	{\footnotesize $\rho=1$}
 	\writefig	4.6	2.3	{\footnotesize $\rho=2$}
 	\writefig	4.55	3.6	{\footnotesize $\rho=3$}
 	\writefig	4.55	4.75	{\footnotesize $\rho=4$}
        \writefig	4.15	0.45	{\footnotesize $0$}
 	\writefig	3.9	1.7	{\footnotesize $2U$}
 	\writefig	3.9	3.0	{\footnotesize $4U$}
 	\writefig	3.9	4.25	{\footnotesize $6U$}
}
\caption{Zero temperature phase diagram for the Bose-Hubbard model. Lobes are
incompressible phases with integer densities.}
\label{phdBH}
\end{figure}

It has been introduced in \cite{FWGF} and despite its simplicity, it has very
interesting phase diagram, see \fig\ref{phdBH}. A phase transition
insulator-superfluid is expected when the hopping coefficient increases. Of
course one would like to have mathematical statements to support this, but the
superfluid phase of interacting particles is hard to
study.\footnote{The best rigorous statements concern the hard-core Bose-Hubbard
model, where off-diagonal long-range order can be proven using
reflection positivity for special value of $\mu$ \cite{DLS}; this constitutes a
beautiful result, although it does not allow to study pure
states.}  On the other hand,
one can tame the insulating phase much more easily. When $t/U$ is small, and
$2U(k-1) < \mu < 2Uk$, it is possible to show that the Gibbs state exists at
low temperature and $\frac1Z \e{-\beta H}$ is close to the projector onto the
configuration $n_x = k$ for all $x \in \bbL$; moreover, the density of the
(quantum) ground state is not only close, but equal to $k$ \cite{BKU}.

\begin{figure}[h]
\epsfxsize=60mm
\centerline{\epsffile{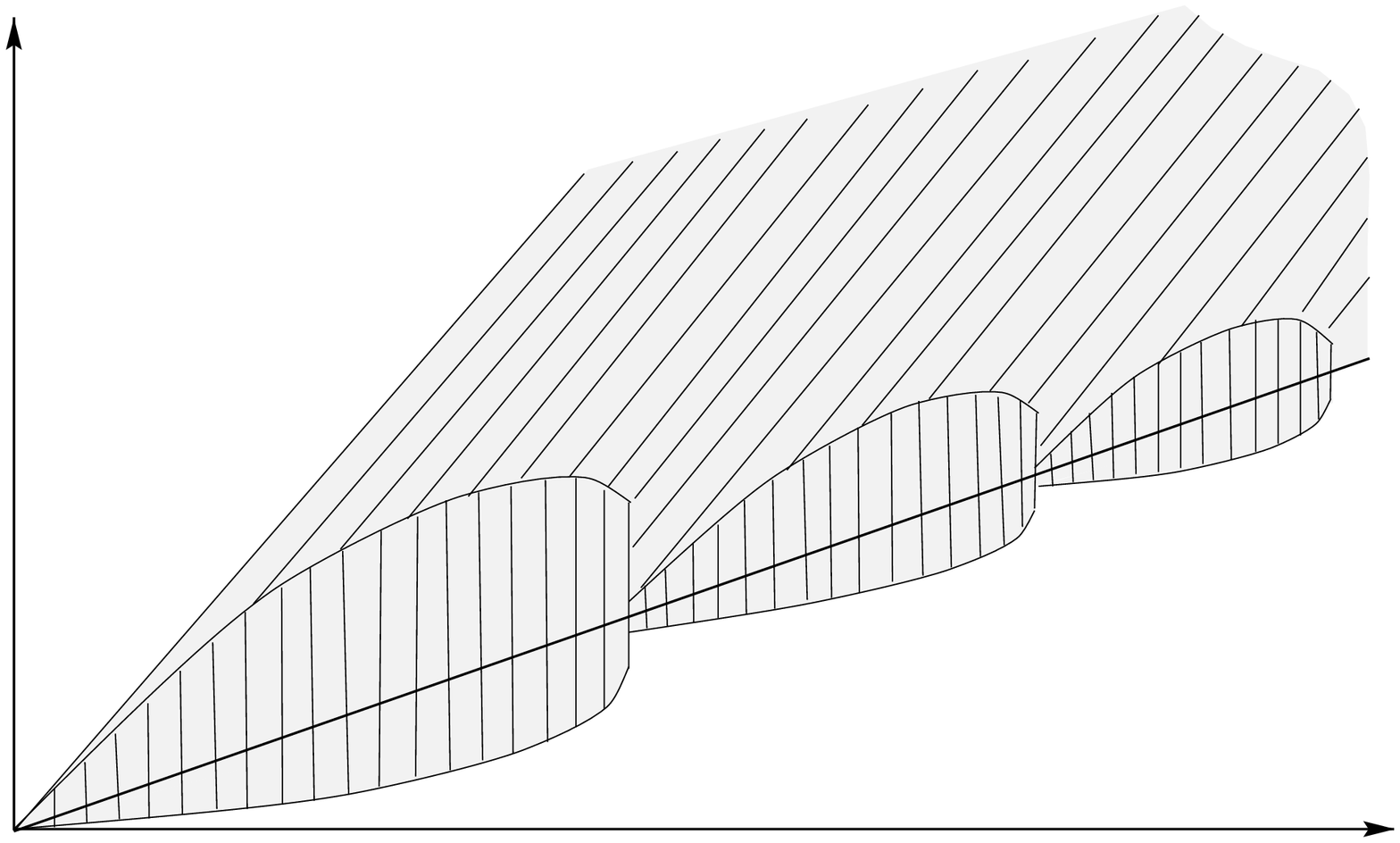}}
\figtext{
        \writefig	10.45	0.4	{$t$}
        \writefig	10.2	2.3	{\small $\mu$}
        \writefig	 4.1	0.4	{\footnotesize $0$}
 	\writefig	 4.5	2.7	{\it\footnotesize uniqueness}
 	\writefig	 4.0	3.9	{$\frac1\beta$}
}
\caption{Analyticity of the free energy and existence of the Gibbs state can be proven on the left
of the grey frontier.}
\label{phdBH2}
\end{figure}

We show here that these phases can be reached from the high temperatures
without phase transition (\fig \ref{phdBH2}). Theorem \ref{thma} does not apply
here, because the single site phase space $\Omega$ is
infinite. Actually, boson systems and unbounded spin systems present some
difficulties at high temperatures, since partition functions diverge at
$\beta=0$. Results for small $\beta$ have been obtained in \cite{PY}. We can prove
analyticity of the free energy and existence of Gibbs state when $\beta t$ is
small, but we are unable to show stability against perturbations, or
against boundary interactions, that do not conserve the total number of
particles (see discussion in Section \ref{secunique}).

Because the phase space has infinite dimension, we need to define the Gibbs state as
a functional over possibly unbounded operators, as for instance $c_x^\dagger$,
or number operators; but not all operators can be considered. In order to define
a suitable class of local operators, let $\tilde N_A$ be the number operator in
$A$ with minimum eigenvalue 1, \ie
\be
\tilde N_A \ket{n_A} = \begin{cases} \ket{n_A} & \text{if $n_x=0$ for all $x \in
A$} \\ (\sum_{x \in A} n_x) \ket{n_A} & \text{otherwise;} \end{cases}
\end{equation}
this operator has an inverse which is defined everywhere. Defining the {\it
boson norm} of a local operator $K$ by
\be
\| K \|^{\text{boson}} \isdefby \sup_{n, n' \in \bbN^{\supp K}} \Bigl| \bra n
\tilde N_{\supp K}^{-\frac12} K \tilde N_{\supp K}^{-\frac12} \ket{n'} \Bigr| ,
\end{equation}
we consider the class $\caK$ of local operators with finite boson norm. It is
not hard to check that $\| \tilde N_{\{ x,y \}}^{-\frac12} c^\dagger_x c_y
\tilde N_{\{ x,y \}}^{-\frac12} \| \leq 1$, and thus $c^\dagger_x c_y \in
\caK$.

\begin{theorem}[Analyticity in the Bose-Hubbard model]\hfill
\label{thmbosons}

There exists a function $\epsilon(U,\mu) > 0$ such that for all
$\beta$ with $\beta t < \epsilon$,
\begin{itemize}
\item the free energy $f(\beta, \mu)$ exists in the thermodynamic limit and is
analytic in $\beta$ and $\mu$;
\item the Gibbs state $\expval\cdot : \caK \to \bbC$, with free or
periodic boundary conditions, converges weakly in the
thermodynamic limit and is exponentially clustering.
\end{itemize}
\end{theorem}

Remark: adding a hard-core condition on the model, \eg by limiting the number of
bosons at a given site to $M < \infty$, Theorem \ref{thmbosons} becomes a
consequence of Theorem \ref{thma} (with moreover uniqueness, and
absence of superfluidity).
Actually, we expect Theorem \ref{thma} to hold when $T$ has finite {\it boson
interaction norm} $\| T \|^{\text{boson}}_c$
\be
\| T \|^{\text{bosons}}_c \isdefby \sup_{x \in \bbL} \sum_{A \ni x} \| \tilde
N_A^{-\frac12} T_A \tilde N_A^{-\frac12} \| \e{c \| A \|} .
\end{equation}
However, we are unable to handle such a general situation; the Bose-Hubbard
model conserves the total number of particles, and this property plays a crucial
technical role.

The present result, together with \cite{BKU}, shows that the free energy is
analytic in a domain that includes low and high temperatures,
corresponding to insulating phase; see \fig
\ref{phdBH2}.

\begin{proof}[Proof of Theorem \ref{thmbosons}]

Expanding $\Tr K \e{-\beta H_\Lambda}$, we obtain \eqref{defrhoK} and
\eqref{expansionK}. We show now exponential decay for $\rho_K(\caA_K)$; similar
(and simpler) considerations lead to exponential decay for $\rho(\caA)$.
Analyticity of the free energy and existence of Gibbs state are then
consequences of Proposition \ref{propclexp} for cluster expansion.

The condition $\cup_{j=1}^m A_j \cup \supp K = \caA_K$ implies
\be
|\caA_K| \leq |\supp K| + m
\end{equation}
(recall that $|A_j|=2$ for all $1 \leq j \leq m$). Then
\bm
|\rho_K(\caA_K)| \leq \e{-c |\caA_K|} \e{c |\supp K|} \e{\beta f_0(\beta, \mu)
|\caA_K|} \\
\sum_{m \geq 0}  t^m \e{cm} \sum_{\neighbours{x_1}{y_1}, \dots,
\neighbours{x_m}{y_m} \subset \caA_K} \int_{0 < \tau_1 < \dots < \tau_m <
\beta} \dd\tau_1 \dots \dd\tau_m \sum_{n_{\caA_K}} \\
\bra{n_{\caA_K}} K \e{-\tau_1 \sum_{x \in \caA_K} V_x^\mu} c_{x_1}^\dagger
c_{y_1} \e{-(\tau_2 - \tau_1) \sum_{x \in \caA_K} V_x^\mu} \dots
c_{x_m}^\dagger c_{y_m} \e{-(\beta - \tau_m) \sum_{x \in \caA_K} V_x^\mu}
\ket{n_{\caA_K}} .
\end{multline}
For given $\ket{n_{\caA_K}}$, let $n_{\caA_K}^1$, \dots, $n_{\caA_K}^m \in
\bbN^{\caA_K}$ such that
\ba
\ket{n_{\caA_K}^m} &\sim c_{x_m}^\dagger c_{y_m} \ket{n_{\caA_K}} \nn\\
\ket{n_{\caA_K}^{m-1}} &\sim c_{x_{m-1}}^\dagger c_{y_{m-1}} \ket{n_{\caA_K}}
\nn\\
&\vdots \nn\\
\ket{n_{\caA_K}^1} &\sim c_{x_1}^\dagger c_{y_1} \ket{n_{\caA_K}^2} . \nn
\end{align}
Then the matrix element in the above equation takes form
\bm
\bra{n_{\caA_K}} K \ket{n_{\caA_K}^1} \e{-\tau_1 \sum_{x \in \caA_K}
\bra{n_{\caA_K}^1} V_x \ket{n_{\caA_K}^1}} \bra{n_{\caA_K}^1} c_{x_1}^\dagger
c_{y_1} \ket{n_{\caA_K}^2} \dots \\
\hfill \dots \bra{n_{\caA_K}^m} c_{x_m}^\dagger
c_{y_m} \ket{n_{\caA_K}} \e{-(\beta-\tau_m) \sum_{x \in \caA_K}
\bra{n_{\caA_K}} V_x \ket{n_{\caA_K}}} \\
\leq \| K \|^{\text{boson}} \bigl( 1 + \sum_{x \in \caA_K} n_x \bigr)
\prod_{j=1}^{m+1} \e{-(\tau_j - \tau_{j-1}) \sum_{x \in \caA_K}
\bra{n_{\caA_K}^j} V_x \ket{n_{\caA_K}^j}} \prod_{j=1}^m (n_{x_j}^{j+1} +
n_{y_j}^{j+1}) ;
\label{borne}
\end{multline}
in the last line we set $\tau_0=0$, $\tau_{m+1} = \beta$ and $n_{\caA_K}^{m+1}
= n_{\caA_K}$. We used the inequality $\bra{n'} c_x^\dagger c_y \ket n \leq n_x
+ n_y$.
Now
\be
\prod_{j=1}^{m+1} \e{-(\tau_j - \tau_{j-1}) \sum_{x \in \caA_K}
\bra{n_{\caA_K}^j} V_x \ket{n_{\caA_K}^j}} \leq \sum_{j=1}^{m+1} \e{-\beta
\sum_{x \in \caA_K} \bra{n_{\caA_K}^j} V_x \ket{n_{\caA_K}^j}} .
\end{equation}
The sum over $n_{\caA_K}$ can be replaced by a sum over $n_{\caA_K}^j$; since
this last bound does not depend any more on $n_{\caA_K}^i$, $i \neq j$, we can
sum over $\neighbours{x_i}{y_i} \subset \caA_K$, and since $\sum_{x \in \caA_K}
n_x^i = \sum_{x \in \caA_K} n_x^j$ for all $i$, the last product of \eqref{borne}
is bounded by
\be
\prod_{i=1}^m \sum_{\neighbours{x_i}{y_i} \subset \caA_K}
(n_{x_i}^{i+1} + n_{y_i}^{i+1}) \leq
\bigl( \chi \sum_{x \in \caA_K} n_x \bigr)^m .
\end{equation}
Collecting these estimates, we get
\bm
|\rho_K(\caA_K)| \leq \e{-c|\caA_K|} \| K \|^{\text{boson}} \e{c|\supp K|}
\e{\beta f_0(\beta,\mu) |\caA_K|} \\
\sum_{m \geq 0} (m+1)\, t^m \e{cm} \frac{\beta^m}{m!} \chi^m \sum_{n_{\caA_K}}
\bigl( 1 + \sum_{x \in \caA_K} n_x \bigr)^{m+1} \e{-\beta
\sum_{x \in \caA_K} \bra{n_{\caA_K}} V_x \ket{n_{\caA_K}}} .
\end{multline}
When $\beta t$ is small, we have $(\beta t \e c \chi)^m \leq (\sqrt{\beta t \e c
\chi})^{m+1}$; therefore
\be
\sum_{m \geq 0} \frac{m+1}{m!} \Bigl[ \bigl( 1 + \sum_{x \in \caA_K} n_x \bigr)
\sqrt{\beta t \e c \chi} \Bigr]^{m+1} \leq \exp\Bigl\{ \sqrt{\beta t \e{c+2}
\chi} \bigl( 1 + \sum_{x \in \caA_K} n_x \bigr) \Bigr\} .
\end{equation}
We obtain finally
\be
|\rho_K(\caA_K)| \leq \e{-c|\caA_K|} \| K \|^{\text{boson}} \e{c|\supp K|}
\Bigl\{ \e{\beta f_0(\beta,\mu)} \sum_{n=0}^\infty \e{n \sqrt{\beta t \e{c+2}
\chi}} \e{-\beta [U(n^2-n) -\mu n]} \Bigr\}^{|\caA_K|} .
\end{equation}
The sum over $n$ is absolutely convergent, so that the quantity
between brackets converge to 1 when $\beta t \to 0$. Having chosen $c$ sufficiently
large to ensure validity of cluster expansion, we see that $\rho_K$ decays
exponentially when $\beta t$ is small enough.

\end{proof}

\subsection*{Acknowledgments} It is a pleasure to thank Volker Bach and Nicolas
Macris for encouragements, and Paul Balmer for discussions. I am also grateful to the referees for
useful comments.

\end{document}